\documentclass[twocolumn,aps,prd,10pt,superscriptaddress,longbibliography,floatfix,citeautoscript]{revtex4-2}

\usepackage{amsmath}
\usepackage{amssymb}
\usepackage{bm}
\usepackage{dcolumn}
\usepackage{glossaries}
\usepackage{graphicx}
\usepackage{multirow}
\usepackage{fancyvrb}
\usepackage{siunitx}
\usepackage[utf8]{inputenc}
\usepackage[usenames,dvipsnames]{xcolor}
\usepackage{hyperref}
\hypersetup{
    pdfnewwindow=true,      % links in new window
    colorlinks=true,        % false: boxed links; true: colored links
    linkcolor=Blue,         % color of internal links
    citecolor=Blue,         % color of links to bibliography
    filecolor=Blue,         % color of file links
    urlcolor=Blue           % color of external links
}

\setacronymstyle{long-short}
\newacronym{gal}{GAL}{graphene antidot lattice}
\newacronym{lsqt}{LSQT}{linear-scaling quantum transport}
\newacronym{md}{MD}{molecular dynamics}
\newacronym{nep}{NEP}{neuroevolution potential}
\newacronym{hnemd}{HNEMD}{homogeneous non-equilibrium molecular dynamics}
\newacronym{tb}{TB}{tight-binding}

\begin{document}

\title{Optimizing thermoelectric performance of graphene antidot lattices via quantum transport and machine-learning molecular dynamics simulations}

\author{Yang Xiao}
%\email{xiaoyanggao1214@163.com}
\affiliation{College of Physical Science and Technology, Bohai University, Jinzhou 121013, P. R. China}

\author{Yuqi Liu}
%\email{lyq2023012027@bhu.edu.cn}
\affiliation{College of Physical Science and Technology, Bohai University, Jinzhou 121013, P. R. China}

\author{Zihan Tan}
%\email{tanzihan2022@163.com}
\affiliation{College of Physical Science and Technology, Bohai University, Jinzhou 121013, P. R. China}

\author{Bohan Zhang}
%\email{zbohan0908@gmail.com}
\affiliation{College of Physical Science and Technology, Bohai University, Jinzhou 121013, P. R. China}

\author{Ke Xu}
%\email{kexu@cuhk.edu.hk}
\affiliation{College of Physical Science and Technology, Bohai University, Jinzhou 121013, P. R. China}

\author{Zheyong~Fan}
\email{brucenju@gmail.com}
\affiliation{College of Physical Science and Technology, Bohai University, Jinzhou 121013, P. R. China}

\author{Shunda Chen}
\email{phychensd@gmail.com}
\affiliation{Department of Civil and Environmental Engineering, George Washington University, Washington, DC 20052, USA}

\author{Shiyun Xiong}
\email{syxiong@gdut.edu.cn}
\affiliation{Guangzhou Key Laboratory of Low-Dimensional Materials and Energy Storage Devices, School of Materials and Energy, Guangdong University of Technology, Guangzhou 510006, China}

\author{Haikuan Dong}
\email{donghaikuan@163.com}
\affiliation{College of Physical Science and Technology, Bohai University, Jinzhou 121013, P. R. China}

\date{\today}

\begin{abstract}
Thermoelectric materials, which can convert waste heat to electricity or be utilized as solid-state coolers, hold promise for sustainable energy applications. 
However, optimizing thermoelectric performance remains a significant challenge due to the complex interplay between electronic and thermal transport properties.
In this work, we systematically optimize $ZT$ in graphene antidot lattices (GALs), nanostructured graphene sheets with periodic nanopores characterized by two geometric parameters: the hexagonal unit cell side length $L$ and the antidot radius $R$.
The lattice thermal conductivity is determined through machine-learned potential-driven molecular dynamics (MD) simulations, while electronic transport properties are computed using linear-scaling quantum transport in combination with MD trajectories based on a bond-length-dependent tight-binding model. 
This method is able to account for electron-phonon scattering, allowing access to diffusive transport in large-scale systems, overcoming limitations of previous methods based on nonequilibrium Green function formalism.
Our results show that the introduction of the antidots effectively decouples lattice and electronic transport and lead to a favorable and significant violation of the Wiedemann-Franz law.
We find that optimal $ZT$ values occur in GALs with intermediate $L$ and $R$, closely correlated with peak power factor values.
Notably, thermoelectric performance peaks near room temperature, with maximal $ZT$ values approaching 2, highlighting GALs as promising candidates for high-performance thermoelectric energy conversion.
\end{abstract}

\maketitle

\section{Introduction}

Thermoelectric materials have attracted significant attention in recent decades due to their ability to directly convert heat into electricity (and vice versa), offering promising applications in energy harvesting, waste heat recovery, and solid-state cooling \cite{bell2008science}.  
These materials demonstrate significant potential for achieving both environmental sustainability and economic benefits across diverse applications \cite{zheng2014rser}.  
In addition to experimental investigations \cite{ma2021mssp}, computational approaches have emerged as a powerful tool for the discovery and design of high-performance thermoelectric materials \cite{zheng2008recent, gorai2017nrm, Kozinsky2021armr, park2025apr}. 

The conversion efficiency of a thermoelectric material at temperature $T$ is quantified by the dimensionless figure of merit $ZT$, defined as 
\begin{equation}
    ZT = \frac{S^2 \sigma T}{\kappa_{\rm l} + \kappa_{\rm e}},
\end{equation}
where $S$ is the Seebeck coefficient, $\sigma$ is the electrical conductivity, and $\kappa_{\rm l}$ and $\kappa_{\rm e}$ are the lattice and electronic thermal conductivities, respectively.  
To maximize $ZT$, a material must simultaneously exhibit a high power factor ($S^2\sigma$) and low total thermal conductivity ($\kappa_{\rm l} + \kappa_{\rm e}$).  

However, these transport parameters are inherently coupled through complex physical mechanisms \cite{Snyder2008nm}: 
(1) The Wiedemann-Franz law \cite{mermin1976} links $\kappa_{\rm e}$ to $\sigma$, limiting their independent optimization. 
(2) Increasing carrier concentration typically enhances $\sigma$ but reduces $S$. 
(3) Phonon scattering strategies to suppress $\kappa_{\rm l}$ may inadvertently affect charge carrier mobility.
This interdependence poses a fundamental challenge in designing materials with high $ZT$ values, necessitating innovative approaches.

Nanostructuring provides a promising approach in decoupling the various transport properties and optimizing $ZT$ \cite{Dresselhaus2007am, Minnich2009ees, szczech2011jmc, YinNatEnerg2023}.
This strategy could significantly alter the transport properties of the underlying bulk material and achieve high $ZT$.
Graphene, in its pristine form, is not a good thermoelectric material due to its zero band gap \cite{sarma2011rmp} and high lattice thermal conductivity \cite{chen2012thermal}.
However, nanostructured graphene systems \cite{Dollfus2015jpcm, zong2020aem} have shown great promise in achieving high $ZT$.
In particular, graphene nanomeshes \cite{bai2010nnt}, or graphene anti-dot lattices  \cite{Ped2008prl}, exhibit a considerable band gap and highly reduced thermal conductivity \cite{fengCarbon2016, wan2020carbon, wei2020ne}, making them promising candidates for useful thermoelectric applications.
Large area smooth-edged graphene nanomesh can be produced by nanosphere lithography \cite{WangSR2013}.
The experimental measurements for thermoelectric transport properties in single- and bi-layer graphene nanomeshes  \cite{Oh2017ne} have demonstrated the proof-of-concept for this promising approach, whereas there remains a large space for further design and optimization.

\begin{figure*}
\centering
\vspace{-0.5cm}
\includegraphics[width=2\columnwidth]{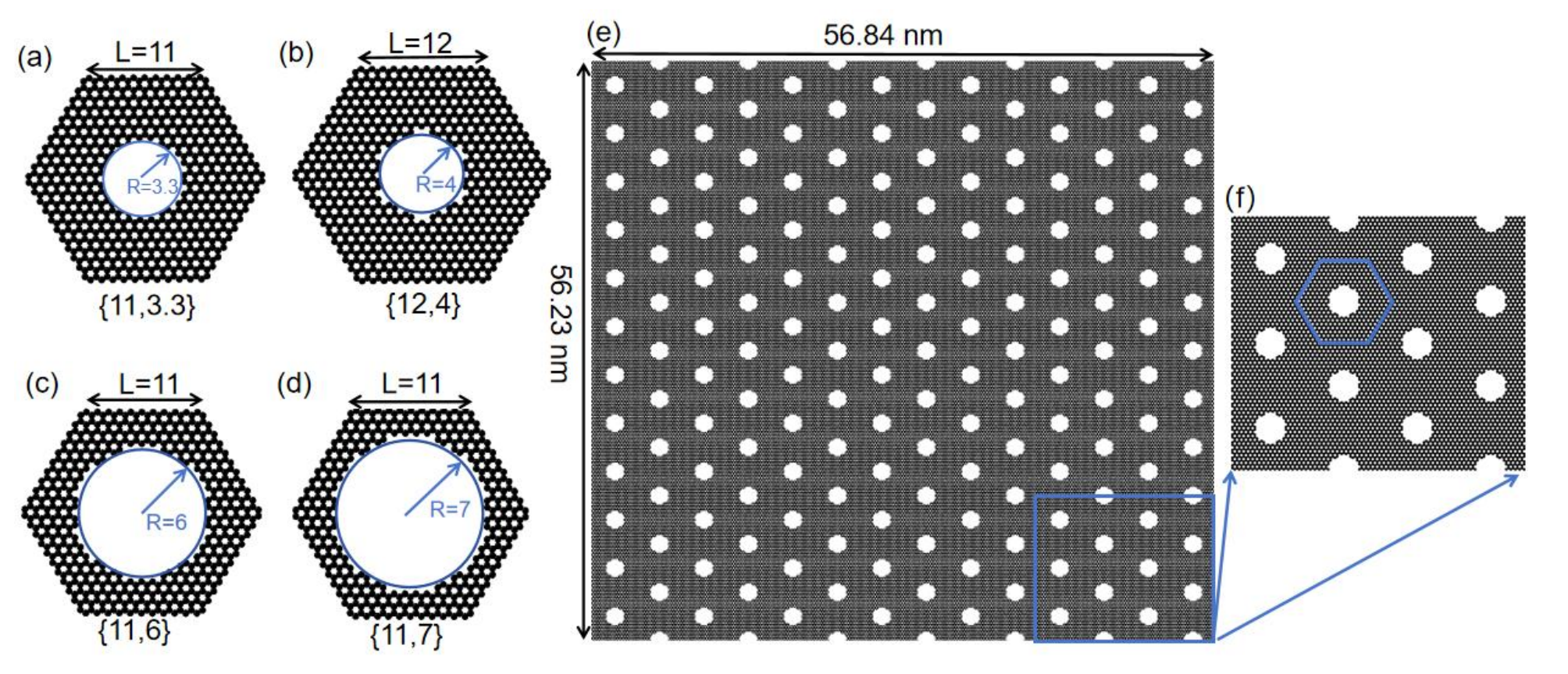}
\caption{Schematic illustration of the simulation models for the graphene antidot lattices (GALs). The hexagonal unit cell of a GAL is represented as $\{L,R\}$, where $L$ is the side length and $R$ is the antidot radius, both in units of  $a=0.246$ nm. (a)-(d) Unit cells for four GALs with different $\{L,R\}$ combinations.
(e) A large super cell for the $\{11,3.3\}$-GAL with dimension $56.84 \times 56.23$ nm$^2$. (f) An illustration of the hexagonal unit cell in the $\{11,3.3\}$-GAL.}
\label{fig:model}
\end{figure*}

Physical insights have been gained by theoretical modeling of thermoelectric transport in \glspl{gal}, using the non-equilibrium Green function (NEGF) method \cite{Gunst2011prb, Karamitaheri2011jap, chang2012prb, GangulySM2019, cmar2020pra}.
The NEGF method, however, has two limitations:
(1) It is computationally efficient only in the ballistic transport regime.
(2) Its cubic scaling in computational cost limits the studied systems to be nanoribbons, usually far from an extended 2D systems. 
These limitations have hindered in-depth investigations of thermoelectric transport in realistically large 2D \glspl{gal}.

In this work, \gls{md} simulations are used not only to obtain the lattice thermal conductivity, but also to generate realistic atomistic configurations, based on which \gls{tb} Hamiltonian for electrons can be constructed using a bond-length dependent \gls{tb} model. 
Consequently, electron-phonon coupling can be effectively captured, offering a way to compute the diffusive electronic transport properties using \gls{lsqt} methodologies \cite{fan2020pr}. 
We systematically investigate the influence of the geometric parameters on $ZT$ of realistic 2D \glspl{gal}, encompassing temperature effects. 
We find that optimal $ZT$ is achieved in GALs with intermediate side length of the hexagonal unit cell and antidot radius. 
At room temperature, the maximal $ZT$ is about 2, highlighting the promise of these GALs for high-performance thermoelectric applications.

\section{Models and Methods}

\subsection{The atomitstic models for graphene antidot lattices}

A \gls{gal} is a graphene sheet with periodically placed nanopores.
The nanopores form a 2D superlattice such as a hexagonal one as shown in Fig.~\ref{fig:model}{(a)}.
Within each hexagon unit cell, there is a hole, or antidot, with a given shape. 
Previous works have suggested that \glspl{gal} with circular antidots exhibit favorable electrical transport properties \cite{Karamitaheri2011jap, GangulySM2019}. 
Therefore, we consider circular antidots in this work, which is also the type proposed in the original work ~\cite{Ped2008prl}.
The center of the antidot is located in the middle of the hexagonal ring of the corresponding pristine graphene lattice.
A hexagonal unit cell with circular antidot is characterized by two geometrical parameters: the side length $L$ of the hexagon and the radius  $R$ of the antidot. 
A \gls{gal} with parameters $L$ and $R$ is then referred to as $\{L, R\}$-\gls{gal}.
In our work, both $L$ and $R$ are in units of the graphene lattice constant $a = 0.246$ nm.
Fig.~\ref{fig:model} (b)-(e) show a few typical unit cells with different sets of $\{L, R\}$.
In order to match the lattice of \gls{gal} to the hexagonal lattice of pristine graphene, $L$ should be an integer and $R$ can be arbitrary. 
It should be noted, however, that for some values of $R$, there exist carbon atoms with only one neighboring atom (two dangling bonds). 
Such atoms are unlikely to exist in a real systems and are therefore removed.
A hexagonal unit cell contains $ 6 L^{2}$ carbon atoms before creating the antidot.
After creating the antidots, the unit cells in Fig.~\ref{fig:model}(b)-(e) contain 642, 750, 468, and 378 atoms, respectively. 
In this work, we have considered a total of 20 \glspl{gal} with the following $\{L,R\}$ values: $\{10, 2\}$, $\{11,2\}$, $\{9,3.3\}$, $\{10,3.3\}$, $\{11,3.3\}$, $\{12,3.3\}$, $\{7,4\}$, $\{8,4\}$, $\{9,4\}$, $\{10,4\}$, $\{11,4\}$, $\{12,4\}$, $\{13,4\}$, $\{9,5\}$, $\{10,5\}$, $\{11,5\}$, $\{12,5\}$, $\{10,6\}$, $\{11,6\}$, and $\{11,7\}$.

In our \gls{md} and \gls{lsqt} simulations, we have used large-scale systems consisting of a large number of unit cells arranged in an almost square cell.
For instance, the simulation system shown in Fig.~\ref{fig:model}{(a)} contains 107856 atoms with in-plane dimensions of $56.8 \times 56.2$ nm$^2$. 
Periodic boundary conditions were applied in the in-plane directions ($x$ and $y$), while the $z$ direction adopts free boundary conditions.
In the transport property calculations, the thickness for all systems in the $z$ direction was chosen as 0.335 nm, which is the equilibrium interlayer distance in graphite.

\subsection{The neuroevolution potential model for general carbon systems }

\gls{md} simulations in this work were driven by the well-established \gls{nep} model for general carbon systems, trained previously by some of the present authors \cite{Fan2024jpcm}.

The \gls{nep} approach \cite{fan2021prb,fan2022jpcm,fan2022jcp, song2024general} generally follows the Behler-Parinello neural network potential methodology \cite{behler2007prl}, but it has a set of more systematically constructed descriptors, both radial and angular. 
The radial descriptors are constructed in terms of Chebyshev polynomials, and the angular descriptors are constructed in terms of the Legendre polynomials, or equivalently, spherical harmonics. 
With a set of well constructed descriptors, \gls{nep} usually uses a neural network architecture with a single hidden layer. 
The neural network maps the descriptor vector $\mathbf{q}^i$ of atom $i$ to its site energy $U_i(\mathbf{q}^i)$:
\begin{equation}
\label{equation:Ui}
U_i = \sum_{\mu=1} ^{N_\mathrm{neu}} w ^{(1)} _{\mu} \tanh\left(\sum_{\nu=1} ^{N_\mathrm{des}} w ^{(0)}_{\mu\nu} q^i_{\nu} - b^{(0)}_{\mu}\right) - b^{(1)}.
\end{equation}
Here, $\tanh(x)$ is the activation function, $w^{(0)}$ are the weight parameters connecting the input layer (with dimension $N_{\rm des}$) and the hidden layer (with dimension $N_{\rm neu}$), $w^{(1)}$ represents the weight parameters connecting the hidden layer and the output layer (the site energy), $b^{(0)}$ represent the bias parameters in the hidden layer, and $b^{(1)}$ is the bias parameter in the output layer. 

The training data were taken from Rowe \textit{et al.} \cite{Rowe2020jcp}. 
The \gls{nep} model has been extensively evaluated and benchmarked against a few other state-of-the-art machine-learned potentials \cite{Ying2025CPR}. It has been shown that this \gls{nep} model achieves reasonable accuracy while has excellent computational speed. 
Recently, this \gls{nep} model has been successfully applied to study heat transport in polycrystalline graphene \cite{Zhou2025jap} and grain boundary structural phase transition in diamond \cite{Lusmall2025}. 
These applications strongly suggest the applicability of the \gls{nep} model in modeling complex carbon materials, include the \glspl{gal} in this work.

All the parameters, including those in the descriptors, are trained through the minimization of a loss function using an evolutionary algorithm. The loss function is defined as a weighted sum of the root-mean-square errors of energy, force, and virial between predictions and reference results in combination with regularization terms.

\subsection{The tight-binding model}

The $p_z$-orbital \gls{tb} model with a hopping parameter of $t_0=-2.7$ eV has been widely used for describing electronic and transport properties of graphene and related nanosctructures \cite{sarma2011rmp}.
However, this \gls{tb} model leads to ballistic transport in graphene systems and structural disorder is needed for diffusive transport. 
In this work, we consider realistic structural disorder due to lattice vibrations.
To this end, we use the well-established Harrison model \cite{Harrison1980} for the hopping parameter.
In this model, the hopping parameter between atoms $i$ and $j$ is inversely proportional to the square of the bond length $r_{ij}$:
\begin{equation}
    H_{ij} = t_0 \left( \frac{r_0}{r_{ij}}\right)^2,
\end{equation}
where $r_0=0.142$ nm is the equilibrium bond length in graphene.
During \gls{md} simulations, the lattice vibrations lead to varying bond lengths and varying hopping integrals. 
This is equivalent to a kind of structural disorder resulting from electron-phonon coupling.
The feasibility of this approach in capturing electron-phonon coupling for electronic transport has been demonstrated in various materials \cite{ishii2009crp,ishii2010prl,ishii2012prb, ortmann2011prb, ciuchi2011prb, fan2017_2dm, fan2018cpc}.

\subsection{The homogeneous nonequilibrium molecular dynamics method for lattice thermal conductivity}

Lattice thermal conductivity can be calculated using \gls{md} simulations.
The \gls{nep} approach has been extensively used to study heat transport using \gls{md} simulations \cite{dong2024jap}.
Here, we use the \gls{hnemd} method \cite{fan2019prb} implemented in the GPUMD package \cite{fan2017cpc}.
In this method, an external driving force $\mathbf{F}_i^{\mathrm{ext}}=\mathbf{F}_{\mathrm{e}} \cdot \mathbf{W}_i$ is applied to each atom to produce a uniform heat flow, where
\begin{equation}
    \mathbf{W}_i = \sum_{j \neq i} \mathbf{r}_{i j} \otimes \frac{\partial U_j}{\partial \mathbf{r}_{j i}} 
\end{equation}
is the per-atom virial tensor for atom $i$.
Here, $\mathbf{F}_{\mathrm{e}}$ is the driving force parameter with the dimension of inverse length and $\mathbf{r}_{i j} \equiv \mathbf{r}_j-\mathbf{r}_i$, $\mathbf{r}_i$ being the position of atom $i$.
The phonon (or lattice) thermal conductivity $\kappa^{\mu\nu}_{\rm l}$ is then calculated according to the following linear response relationship
\begin{equation}
\label{equation:hnemd_kappa}
\langle J_{\mu} \rangle = T\Omega \sum_{\nu} \kappa^{\mu\nu}_{\rm l} F^{\nu}_{\rm e}.
\end{equation}
Here $T$ is the system temperature, $\Omega$ is the system volume, and $\langle J_{\mu} \rangle$ is the ensemble average of the $\mu$-component of the heat current \cite{fan2015prb}
\begin{equation}
    \mathbf{J} = \sum_i \mathbf{W}_i \cdot \mathbf{v}_i,
\end{equation}
where $\mathbf{v}_i$ is the velocity of atom $i$.
In this work, we will not consider off-diagonal elements of the thermal conductivity tensor. 
In this case, we can bring Eq.~(\ref{equation:hnemd_kappa}) into a scalar form:
\begin{equation}
    \kappa_{\rm l} = \frac{\langle J \rangle}{T\Omega F_{\rm e}}.
\end{equation}
The lattice thermal conductivity can also be resolved with respect to phonon frequency:
\begin{equation}
    \kappa_{\rm l} = \int \frac{\text{d} \omega}{2\pi}\kappa_{\rm l}(\omega),
\end{equation}
\begin{equation}
    \kappa_{\rm l}(\omega) = \frac{2 \tilde{K}(\omega)}{T\Omega F_{\rm e}},
\end{equation}
where $\tilde{K}(\omega)$ is the Fourier transform of the virial-velocity time-correlation function:
\begin{equation}
    \mathbf{K}(t) = \sum_i \langle \mathbf{W}_i(0) \cdot \mathbf{v}_i(t) \rangle.
\end{equation}

In the \gls{hnemd} simulations, the system was first equilibrated for 0.1 ns in the isothermal-isobaric ensemble at the target temperature and pressure. 
In all simulations, the target pressure was set to zero. 
Subsequently, the system was switched to the isothermal-isochoric ensemble and the external driving force with the strength ${F}_{\rm e}$= $1 \times 10^{-4}$ \AA$^{-1}$ was applied.
We performed 4 independent simulations in both the $x$ and the $y$ directions and averaged the results over the 8 simulations because the \glspl{gal} we considered are essentially isotropic in the in-plane directions.
A time step of 1 fs was used in all \gls{md} simulations.

\subsection{The linear scaling quantum transport approach for thermoelectric transport}

For electronic transport properties, we use the 
\gls{lsqt} method \cite{fan2020pr} that is capable of simulating large systems with linear scaling computational cost. 
The integration of \gls{lsqt} with \gls{nep}-driven \gls{md} for capturing electron-phonon coupling has been proposed recently \cite{Fan2024jpcm} and a proof-of-concept application to thermoelectric transport in \glspl{gal} has been demonstrated. 
In this method, the fundamental quantity is the running differential conductivity $\Sigma(E,t)$ as a function of Fermi energy $E$ and correlation time $t$, which can be expressed as the following Green-Kubo integral:
\begin{equation}
\label{equation:sigma}
    \Sigma(E,t)=\frac{2e^2}{\Omega} \int_0^{t} \mathrm{Tr} \left[\delta (E-\hat{H}) \mathrm{Re} (\hat{V}\hat{V}(\tau)) \right] \text{d}\tau,
\end{equation}
where $e$ is the elementary charge, $\Omega$ is the system volume, $\hat{H}$ is the electron Hamiltonian operator, $\delta(E-\hat{H})$ is the energy resolution operator, $\hat{V}$ is the velocity operator, and $\hat{V}(\tau)=e^{i\hat{H}\tau} \hat{V} e^{-i\hat{H}\tau}$ is the time-evolved velocity operator.
The factor of 2 represents spin degeneracy.
During the \gls{md} simulation, the electron Hamiltonian is updated at every step according to the atom positions. 

 \begin{figure}
\centering
\includegraphics[width=1\columnwidth]{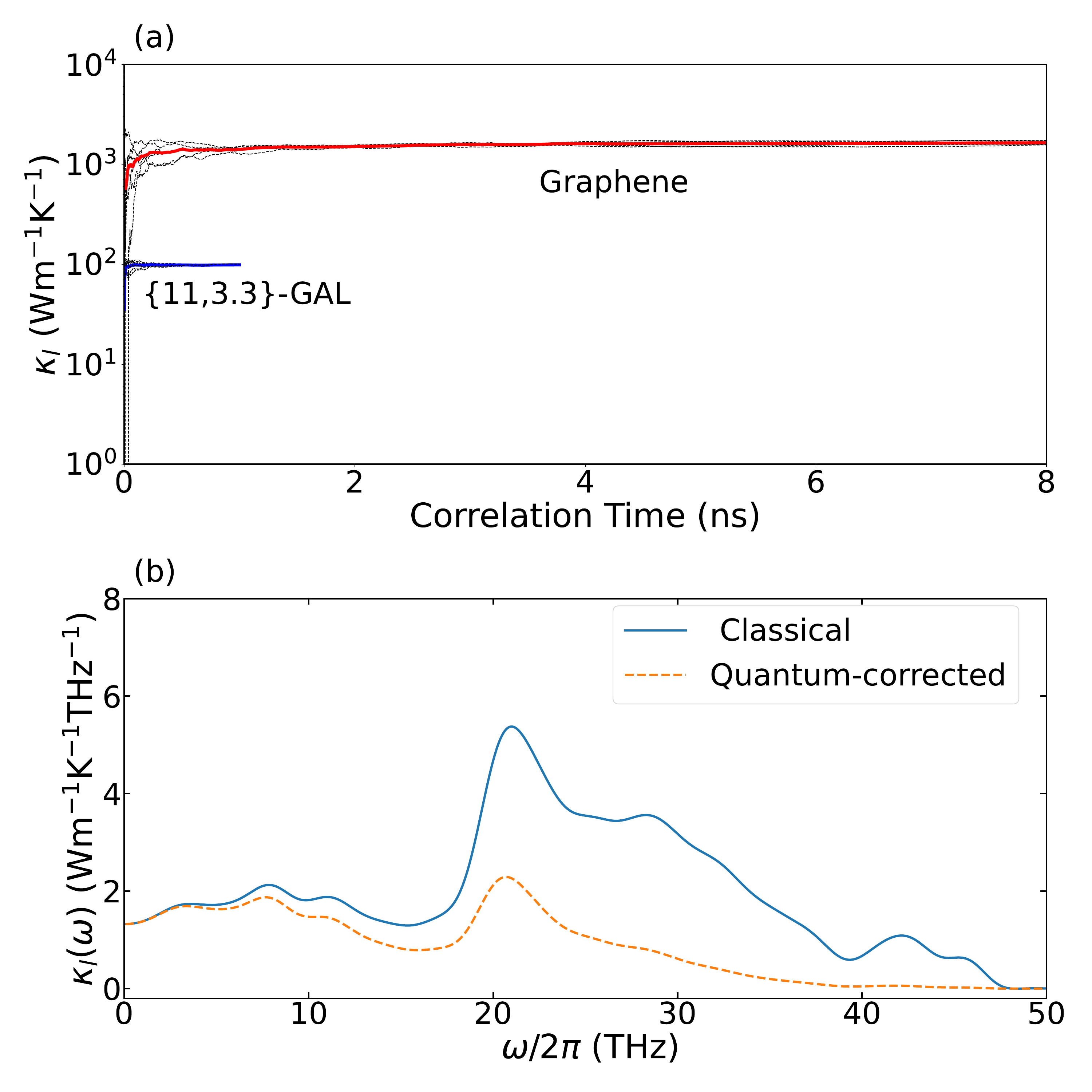}
\caption{(a) Lattice thermal conductivity $\kappa_{\mathrm{l}}$ of pristine graphene and the $\{11,3.3\}$-\gls{gal} as a function of \gls{hnemd} correlation time at \SI{300}{K}. The thin lines represent results from eight independent simulations and the thick lines indicate their averages.  (b) Classical and quantum-corrected spectral thermal conductivity of the $\{11,3.3\}$-\gls{gal} as a function of the phonon frequency $\omega/2\pi$.}
\label{fig:kappa}
\end{figure}

\begin{figure*}[ht]
\centering
\includegraphics[width=2\columnwidth]{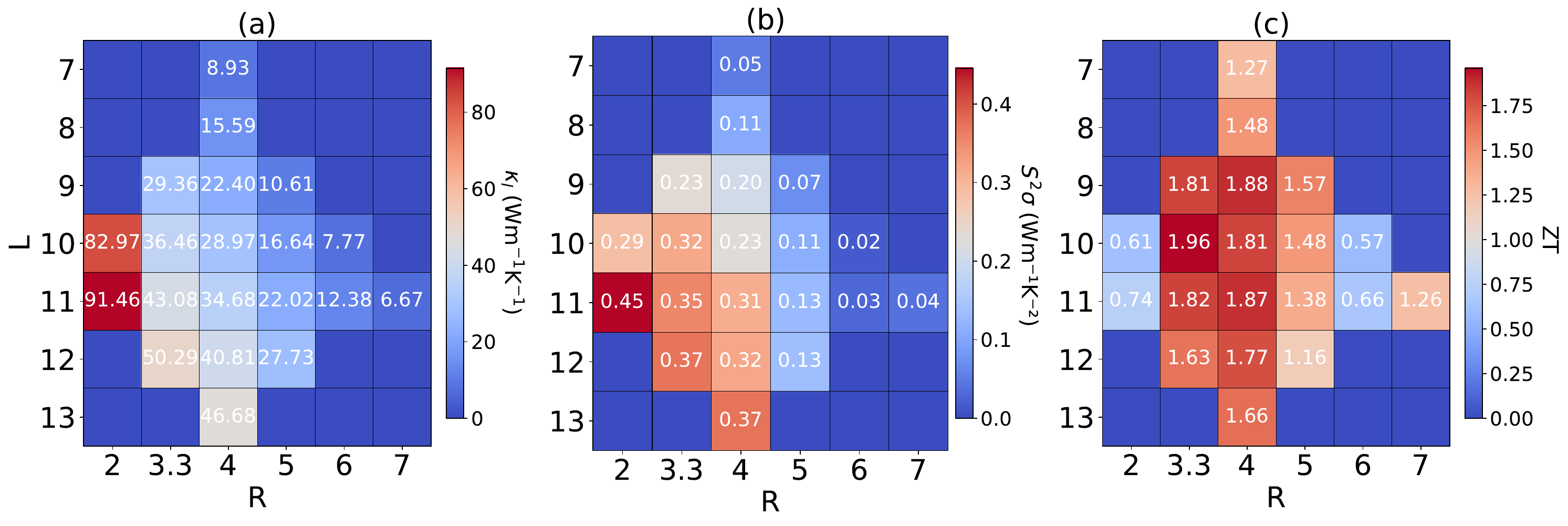}
\caption{Thermoelectric transport properties of all the graphene antidot lattices (GALs) at 300 K considered in this work. (a) Lattice thermal conductivity ($\kappa_{\rm l}$), (b) power factor ($S^2\sigma$), and (c) figure of merit ($ZT$) as functions of the side length ($L$) of the hexagonal unit cell and the antidot radius ($R$). Numerical scale bars and values are provided.
GALs represented by blue blocks without numbers were not calculated.}
\label{fig:kappa-PF-ZT}
\end{figure*}

The time-dependent electrical conductivity $\Sigma(E,t)$ converges in the diffusive transport regime, and one can obtain the transport distribution function \cite{Mahan1996pnas,fan2011jap,Zhou2011prl,Jeong2012jap,Maassen2021prb,ding2023npjcm} $\Sigma(E)$ as the time-converged $\Sigma(E,t)$.
From the transport distribution function, we then calculated the transport coefficients for a range of chemical potential $\mu$. 
We first define the following functionals ($n=0,1,2$) of the transport distribution function:
\begin{equation}
 X_n(\mu,T) = \int_{-\infty}^{\infty} \left[-\frac{\partial f(E,\mu,T)}{\partial E}\right] E^n\Sigma(E) \text{d}E,
\end{equation}
where 
\begin{equation}
f(E,\mu,T) = \frac{1}{\exp\left(\frac{E-\mu}{k_{\rm B} T} \right)+1}
\end{equation} 
is the Fermi-Dirac distribution. 
The electrical conductivity $\sigma(\mu,T)$, Seebeck coefficient $S(\mu,T)$, and electronic thermal conductivity $\kappa_{\rm e}(\mu,T)$ can be expressed as
\begin{equation}
\sigma(\mu,T) = X_0(\mu,T),
\end{equation}
\begin{equation}
S(\mu,T) = -\frac{1}{eT} \left[ \frac{X_1(\mu,T)}{X_0(\mu,T)}-\mu\right],
\end{equation}
\begin{equation}
\kappa_{\rm e}(\mu,T) = \frac{1}{e^2T} \left[X_2(\mu,T) - \frac{X_1^2(\mu,T)}{X_0(\mu,T)}\right].
\end{equation}
Based on these transport coefficients and the phonon thermal conductivity $\kappa_{\rm l}(T)$, we can define the thermoelectric figure of merit as 
\begin{equation}
 ZT(\mu,T) = \frac{S^2(\mu,T) \sigma(\mu,T)} { \kappa_{\rm l}(T) + \kappa_{\rm e}(\mu,T) } T.
\end{equation}

Similarly to the case of thermal transport, we performed at least 4 independent simulations in both the $x$ and the $y$ directions and averaged the results because the \glspl{gal} we considered are essentially isotropic in the in-plane directions.
Moreover, the \gls{tb} model we used has particle-hole symmetry and we have symmetrized the transport distribution function with respect to the Fermi energy. 

\begin{figure}[ht]
\centering
\includegraphics[width=0.5\textwidth]{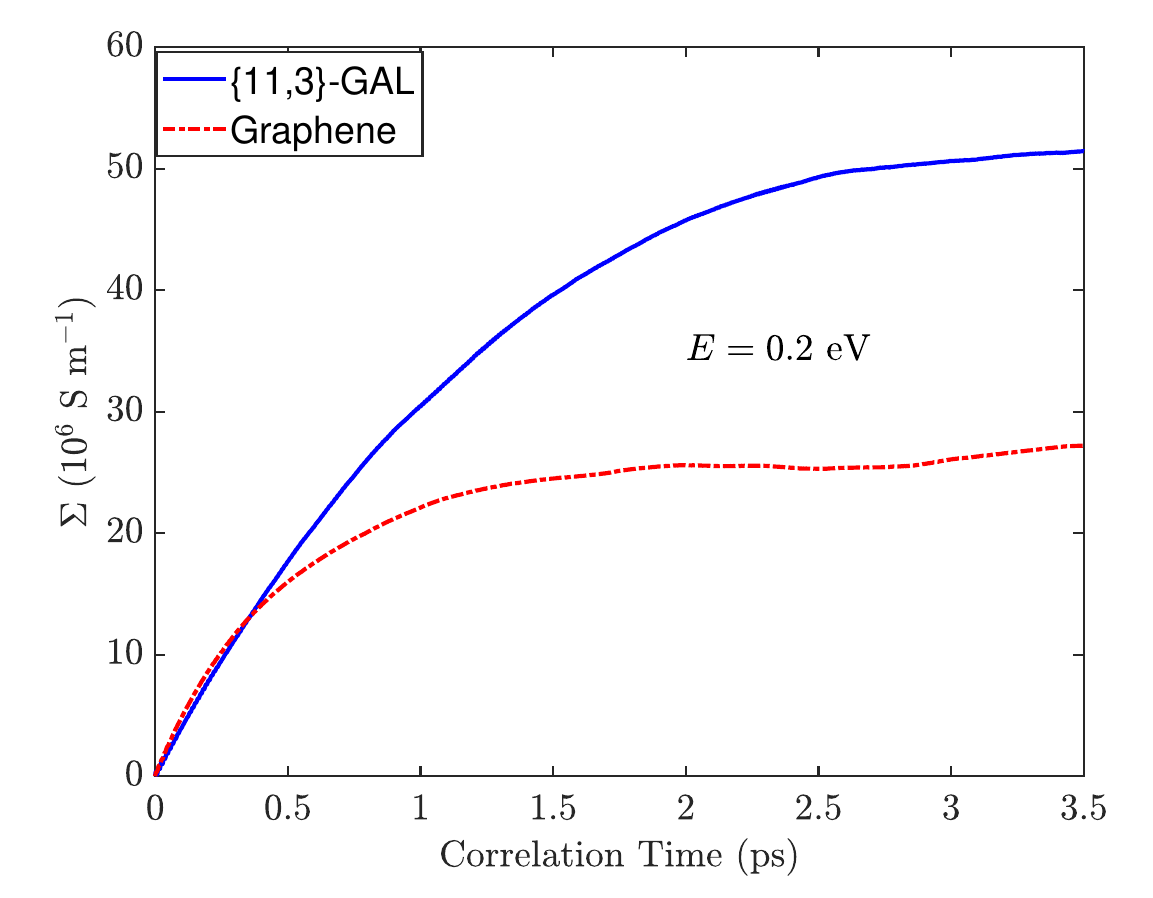}
\caption{Running electrical conductivity. The running electrical conductivity $\Sigma(E,t)$ as a function of correlation time for $\{11,3.3\}$-\gls{gal} (solid line) and pristine graphene (dashed line), both evaluated at the Fermi energy $E=0.2$ eV.}
\label{fig:electronic_time}
\end{figure}

\section{Results and Discussion}

\subsection{Lattice thermal conductivity}

We first examine the lattice thermal conductivity.
To illustrate the process of determining $\kappa_{\rm l}$, we use the $\{11,3.3\}$-\gls{gal} as a representative example.
Fig.~\ref{fig:kappa}{(a)} shows the cumulative average of the lattice thermal conductivity for pristine graphene and the $\{11,3.3\}$-\gls{gal} at 300 K, plotted as a function of correlation time in the \gls{hnemd} simulation. 
We observe that the lattice thermal conductivity converges well over time. 
Overall, the lattice thermal conductivity is reduced by more than an order of magnitude compared to pristine graphene, similar to the reduction seen in nanostructured silicon membranes \cite{Oliveira2025acsanm} and nanowires \cite{xiong2016prl}.
This substantial reduction of $\kappa_{\rm l}$ is a key prerequisite for achieving a considerably large $ZT$ in \glspl{gal}.

However, the lattice thermal conductivity calculated in this way is purely classical.
Due to the high Debye temperature of graphene, quantum effects are strong and a proper quantum correction is needed. 
Because the antidots can be considered as defects in graphene, the phonon-defect scattering dominates and the quantum-correction scheme based on the spectral thermal conductivity \cite{GU2021jap, wang2023prb} can be applied. 
In this method, the quantum spectral thermal conductivity is obtained by multiplying the classical spectral thermal conductivity by the factor $\frac{x^2 e^x}{(e^x - 1)^2}$, where $x=\hbar \omega / k_{\mathrm{B}} T$, $\omega$ is the phonon frequency, $\hbar$ is the reduced Planck constant, and $k_{\mathrm{B}}$ is the Boltzmann constant. 
An example is shown in Fig.~\ref{fig:kappa}{(b)} for the $\{11,3.3\}$-\gls{gal}.
For this structure, $\kappa_{\rm l}$ is reduced from 98.91 to 43.08 W m$^{-1}$ K$^{-1}$.
We have similarly applied the quantum correction to all other \gls{gal} structures (see Supplemental Material Figs. S1-S19).

\begin{figure}[ht]
\centering
\includegraphics[width=0.5\textwidth]{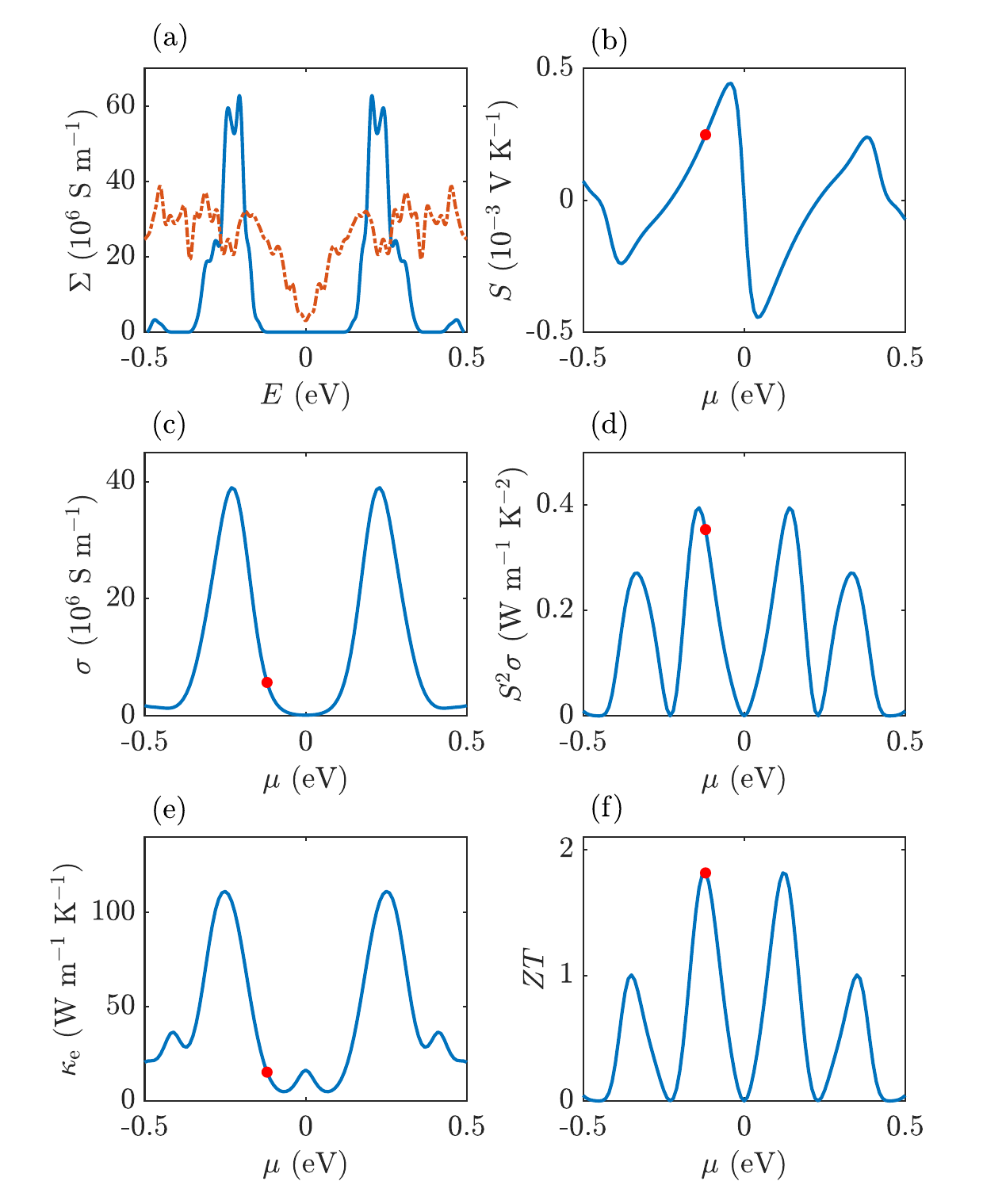}
\caption{Electronic and transport properties of the $\{11,3.3\}$-GAL at 300 K. (a) Transport distribution function ($\Sigma$) as a function of the Fermi energy $E$, (b) Seebeck coefficient ($S$), (c) electrical conductivity ($\sigma$), (d) power factor ($S^2\sigma$), (e) electronic thermal conductivity ($\kappa_{\rm e}$), and (f) figure of merit ($ZT$) as a function of the chemical potential $\mu$.
The dots in (b)-(f) indicate the chemical potential at which $ZT$ is maximal. The dashed line in (a) shows the transport distribution function for pristine graphene.
}
\label{fig:electronic}
\end{figure}

The quantum-corrected lattice thermal conductivity $\kappa_{\rm l}$ for various $L$ and $R$ values are shown in Fig.~\ref{fig:kappa-PF-ZT}(a). 
It is clear to see that $\kappa_{\rm l}$ decreases with decreasing $L$ and increasing $R$.
This is understandable, as smaller $L$ and larger $R$ lead to less conducting channel and more defect scattering. 
Notably, the increase of $R$ from 2 to 3.3 results in a more than two fold decrease of $\kappa_{\rm l}$.

\subsection{Electronic transport properties and figure of merit}

We next examine the electronic transport properties. 
Figure~\ref{fig:electronic_time} illustrates the running electrical conductivity, $\Sigma(E,t)$, calculated using Eq.~(\ref{equation:sigma}) for $\{11,3.3\}$-\gls{gal} and pristine graphene, both evaluated at the Fermi energy of $E=0.2$ eV. 
Interestingly, the $\{11,3.3\}$-\gls{gal} achieves a higher converged electrical conductivity, in sharp contrast to the trend observed for lattice thermal conductivity. 
This suggests that the antidots in the \gls{gal} effectively decouple lattice and electronic transport properties, a desirable effect for achieving high thermoelectric performance.

Figure~\ref{fig:electronic} shows the transport distribution function as a function of the Fermi energy $E$ and the thermoelectric transport properties as a function of the chemical potential $\mu$, all evaluated at the temperature of $T=300$ K. 
Results for all the other \glspl{gal} are presented in Figs. S23-S41.
From the transport distribution function [Fig.~\ref{fig:electronic}(a)], a clear band gap in the \gls{gal} is observed. 
The emergence of this band gap is a manifest of quantum confinement effects, which also give rise to pronounced peaks in the transport distribution function.
These peaked features induced by the antidots contribute to the considerably enhanced Seebeck coefficients [Fig.~\ref{fig:electronic}(b)], while the electrical conductivities [Fig.~\ref{fig:electronic}(c)] remain high.
As a result, high power factors $S^2\sigma$ [Fig.~\ref{fig:electronic}(d)] are achieved.
The electronic thermal conductivity $\kappa_{\rm e}$ is roughly proportional to the electrical conductivity, in accordance with the Wiedemann-Franz law \cite{mermin1976}.
Combining all these electronic transport properties and the lattice thermal conductivity determined above, we obtain the figure of merit $ZT$, as shown in Fig.~\ref{fig:electronic}(f).

All the electronic transport properties are functions of the chemical potential $\mu$, which can be tuned by doping or gating. 
From Fig.~\ref{fig:electronic}(f), we can identify the chemical potential at which the maximal $ZT$ is achieved. 
This chemical potential does not give either the maximal $S$ or the maximal $\sigma$, but roughly corresponds to the maximal power factor $S^2\sigma$.
This trend is valid for all the \glspl{gal} studied in this work (see the Supplemental Material Figs. S23-S41).
This highlights the critical role of the power factor in determining the thermoelectric figure of merit.

The power factors for all the \glspl{gal} we considered are shown in Fig.~\ref{fig:kappa-PF-ZT}(b).
They show qualitatively similar dependence on the geometric parameters $L$ and $R$ as the lattice thermal conductivity, decreasing with decreasing $L$ and increasing $R$.
However, when $R$ is increased from 2 to 3.3, the power factor is not significantly reduced, which is in contrast with the lattice thermal conductivity.
For larger $R$ values, the power factor is very small, mainly due to small electrical conductivity caused by localized states around the edge of the anti-dots \cite{vanPRB2009}. 
The quantitatively different behaviors of the lattice thermal conductivity and the power factor lead to a sharp increase of $ZT$ when $R$ is increased from 2.0 to 3.3, as shown in Fig.~\ref{fig:kappa-PF-ZT}(c).
Overall, high $ZT$ values are obtained for intermediate $L$ (9 to 12) and R (3.4 and 4.0) values.
The maximum $ZT=1.96$ is obtained in $\{10,3.3\}$-\gls{gal}.

\begin{figure}[ht]
\centering
\includegraphics[width=0.5\textwidth]{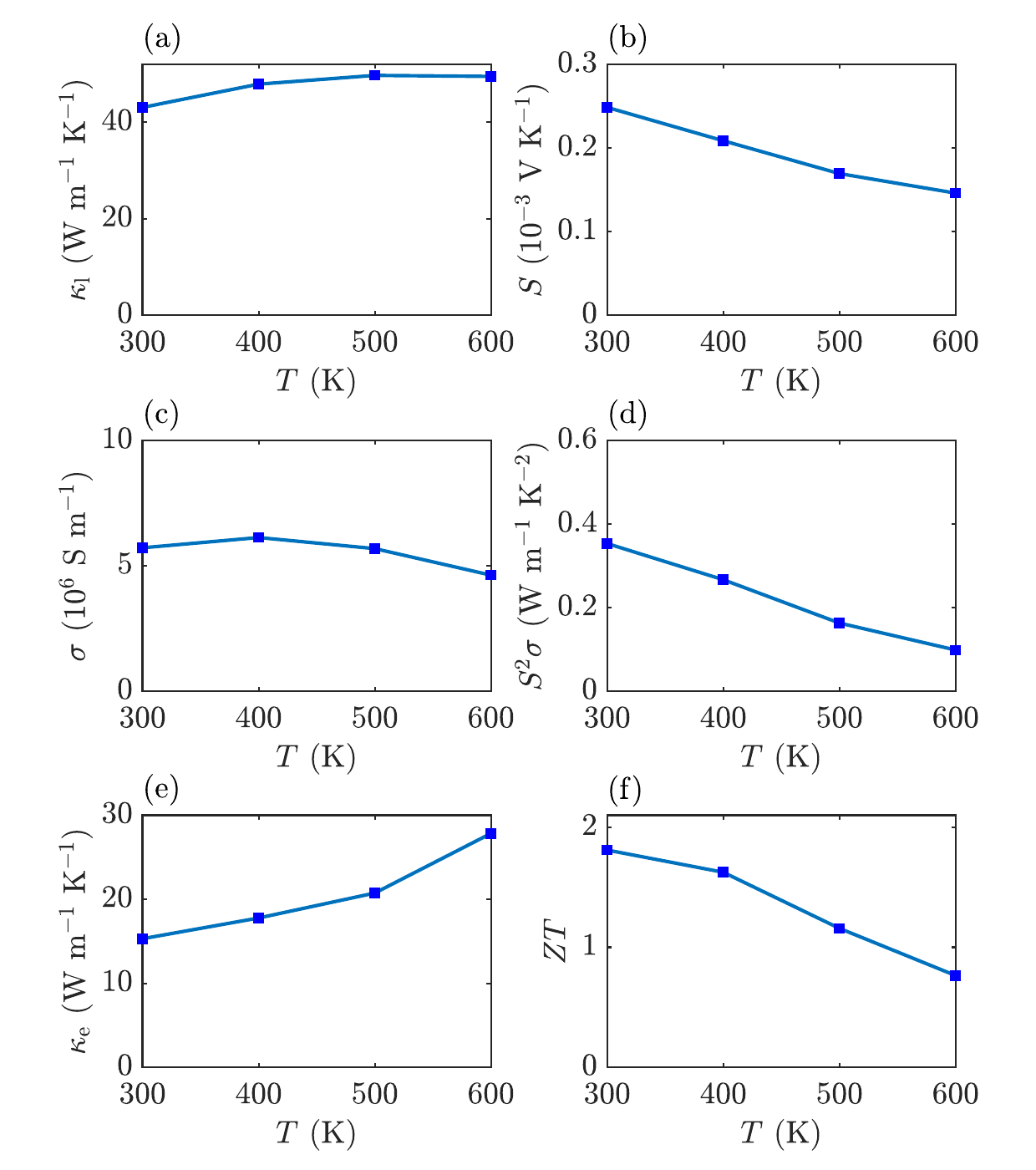}
\caption{Temperature dependence of the thermoelectric transport properties for the $\{11,3.3\}$-GAL. (a) Lattice thermal conductivity ($\kappa_{\rm l}$), (b) Seebeck coefficient ($S$), (c) electrical conductivity ($\sigma$), (d) power factor ($S^2\sigma$), (e) electronic thermal conductivity ($\kappa_{\rm e}$), and (f) thermoelectric figure of merit ($ZT$) for the $\{11,3.3\}$-GAL as functions of temperature $T$. All quantities are evaluated at the chemical potential that maximizes $ZT$.}
\label{fig:temperature}
\end{figure}

\subsection{Temperature dependence and violation of Wiedemann-Franz law}

The above discussions focus on the room temperature.
We next examine the influence of temperature on the thermoelectric performance of \glspl{gal}, taking the $\{11,3.3\}$-\gls{gal} as an example.
Detailed results for lattice thermal conductivity and electronic transport properties are presented in Figs. S20-S22 and Figs. S42-S44.
Figure~\ref{fig:temperature} shows its $\kappa_{\rm l}$ and electronic transport properties from 300 to 600 K.
With increasing temperature, $\kappa_{\rm l}$ [Fig.~\ref{fig:temperature}(a)] increases essentially monotonically due to the increase of the activated phonon modes with increasing temperature, which dominates over phonon scattering within this range of temperature.
The Seebeck coefficient $S$ [Fig.~\ref{fig:temperature}(b)] decreases with increasing temperature due to the smearing of the Fermi-Dirac distribution over the zero-temperature transport distribution function.
The electrical conductivity $\sigma$ [Fig.~\ref{fig:temperature}(c)] also decreases overall with increasing temperature, which can be understood based on the increased electron-phonon scattering.
Combining the behaviors of $S$ and $\sigma$ leads to a strongly decreasing power factor [Fig.~\ref{fig:temperature}(d)] with increasing temperature.
$\kappa_{\rm e}$ shows an opposite trend to $\sigma$ [Fig.~\ref{fig:temperature}(e)], in accordance with the Wiedemann-Franz law, $\kappa_{\rm e} \propto \sigma T$.
Even though there is an extra factor of $T$ in $ZT$, $ZT$ still exhibits a monotonically decreasing trend with increasing temperature. 
Therefore, the \glspl{gal} demonstrate superior thermoelectric performance at room temperature. 
This makes them promising candidates for thermoelectric cooling applications, while less appealing for electricity generation applications at high temperature.

\begin{figure}[ht]
\centering
\includegraphics[width=0.5\textwidth]{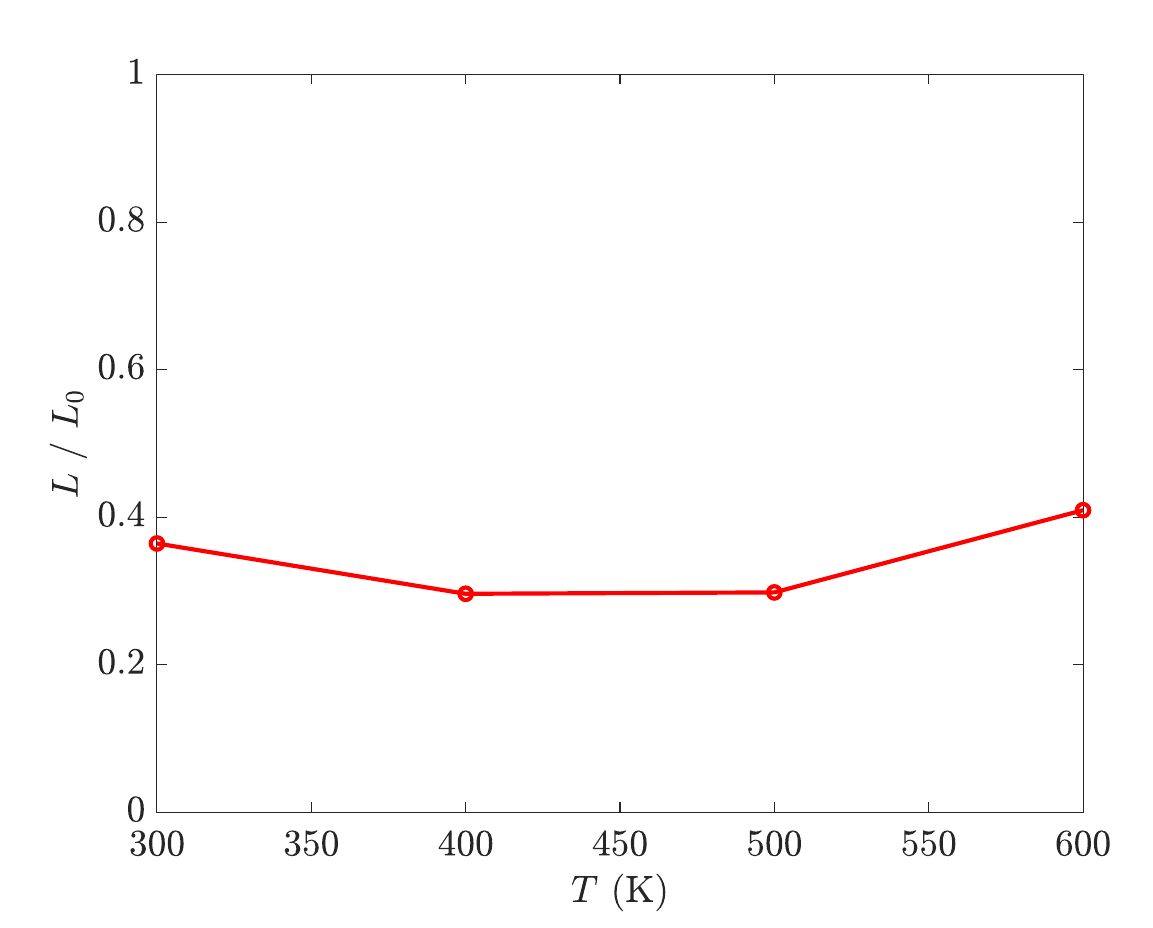}
\caption{Violation of the Wiedemann-Franz law. The ratio $L/L_0$ between the calculated effective Lorenz number $L=\kappa_{\rm e}/\sigma T$ and the theoretical value, $L_0=\pi^2k_{\rm B}^2/3e^2$, predicted by Wiedemann-Franz law, for the $\{11,3.3\}$-GAL as a function of the temperature $T$.}
\label{fig:lorenz}
\end{figure}

Even though $\kappa_{\rm e} \propto \sigma T$, the proportionality coefficient, known as the effective Lorenz number $L=\kappa_{\rm e}/ \sigma T$, turns out to be much smaller than the theoretical one $L_0=\pi^2k_{\rm B}^2/3e^2$, see Fig.~\ref{fig:lorenz}.
The ratio $L/L_0$ is between 0.3 and 0.4.
This amounts to a strong violation of the Wiedemann-Franz law, and is one of the crucial mechanisms  for the high $ZT$ values in the \glspl{gal}.
This large violation of the Wiedemann-Franz law originates from the clearly peaked transport distribution function [Fig.~\ref{fig:electronic}(a)] because the electronic thermal conductivity is proportional to the variance of the transport distribution function \cite{fan2011jap}.

\section{Summary and conclusions}

In summary, we have conducted a comprehensive study on optimizing thermoelectric performance in \glspl{gal} using an approach combining machine-learning \gls{md} and \gls{lsqt} simulations.
The lattice thermal conductivity is calculated through \gls{md} simulations driven by the highly accurate and efficient neuroevolution potential.
Electronic transport properties, including electrical conductivity, Seebeck coefficient, and electronic thermal conductivity, are computed using \gls{lsqt} in combination with MD, based on a bond-length-dependent tight-binding model. 
This approach enables access to diffusive transport properties in large-area systems that have not been feasible for previous approaches based on nonequilibrium Green function formalism.

We systematically investigate the influence of the geometric parameters on $ZT$ of realistic 2D \glspl{gal}, including temperature effects. 
We find that optimal $ZT$ values are achieved in GALs with intermediate $L$ and $R$ and are highly correlated with the optimal power factor values.
The presence of antidots effectively decouples lattice and electronic transports, resulting in a favorable and significant violation of the Wiedemann-Franz law, which enhances thermoelectric performance.
Remarkably, at room temperature, the maximal $ZT$ approaches 2, making these GALs promising candidates for efficient thermoelectric energy conversion.

Further improvement in $ZT$ may be achieved by incorporating resonant structures into the \glspl{gal} design, which may further suppress lattice thermal conductivity by blocking low-frequency phonons \cite{WangNanoscale2021, WuPRApplied2024}.

\begin{acknowledgments}
This work was supported by the National Science and Technology Advanced Materials Major Program of China (No. 2024ZD0606900). 
SX acknowledges the support of the National Natural Science Foundation of China (Grant No. 12174276), the Basic and Applied Basic Research Foundation of Guangdong Province (Grant No. 2024A1515010521).
HD is supported by the Science Foundation from Education Department of Liaoning Province (No. JYTMS20231613) and the Doctoral start-up Fund of Bohai University (No. 0523bs008).
\end{acknowledgments}

\end{document}